\documentclass[prl,aps,amsfonts,amsmath,amssymb,nofootinbib,twocolumn,superscriptaddress]{revtex4}

\usepackage{graphicx}% Include figure files
\usepackage{bm}% bold math
\usepackage{hyperref}
\usepackage{IEEEtrantools}
\usepackage{xcolor}

\usepackage{natbib}
\usepackage{float}

\restylefloat{table}

\begin{document}

	\title{Evidence for highly damped Higgs mode in infinite-layer nickelates}
	
	\affiliation{Ames National Laboratory, Ames, IA 50011 USA}
	
	\affiliation{Stanford Institute for Materials and Energy Sciences, SLAC National Accelerator Laboratory, Menlo Park, CA 94025, USA}
	
	\affiliation{Department of Physics, Stanford University, Stanford, CA 94305, USA}
	
	\affiliation{Department of Applied Physics, Stanford University, Stanford, CA 94305, USA}
	
	\affiliation{Department of Physics, University of Alabama at Birmingham, Birmingham, AL 35294-1170, USA.}

	\affiliation{Department of Physics and Astronomy, Iowa State University, Ames, Iowa 50011, USA.}
	
	\author{Bing Cheng}
	\email{bcheng2@ameslab.gov}
	\affiliation{Ames National Laboratory, Ames, IA 50011 USA}

	\author{Di Cheng}
	\affiliation{Ames National Laboratory, Ames, IA 50011 USA}

	\author{Kyuho Lee}
	\affiliation{Stanford Institute for Materials and Energy Sciences, SLAC National Accelerator Laboratory, Menlo Park, CA 94025, USA}
	\affiliation{Department of Physics, Stanford University, Stanford, CA 94305, USA}

	\author{Martin Mootz}
	\affiliation{Ames National Laboratory, Ames, IA 50011 USA}
	\affiliation{Department of Physics and Astronomy, Iowa State University, Ames, Iowa 50011, USA.}
	
	\author{Chuankun Huang}
	\affiliation{Ames National Laboratory, Ames, IA 50011 USA}
	\affiliation{Department of Physics and Astronomy, Iowa State University, Ames, Iowa 50011, USA.}

	\author{Liang Luo}
	\affiliation{Ames National Laboratory, Ames, IA 50011 USA}
	
	\author{Zhuoyu Chen}
	\affiliation{Stanford Institute for Materials and Energy Sciences, SLAC National Accelerator Laboratory, Menlo Park, CA 94025, USA}
	\affiliation{Department of Applied Physics, Stanford University, Stanford, CA 94305, USA}
	
	\author{Yonghun Lee}
	\affiliation{Stanford Institute for Materials and Energy Sciences, SLAC National Accelerator Laboratory, Menlo Park, CA 94025, USA}
	\affiliation{Department of Applied Physics, Stanford University, Stanford, CA 94305, USA}

	\author{Bai Yang Wang}
	\affiliation{Stanford Institute for Materials and Energy Sciences, SLAC National Accelerator Laboratory, Menlo Park, CA 94025, USA}
	\affiliation{Department of Physics, Stanford University, Stanford, CA 94305, USA}

	\author{Ilias E. Perakis}
	\affiliation{Department of Physics, University of Alabama at Birmingham, Birmingham, AL 35294-1170, USA.}
	
	\author{Zhi-Xun Shen}
	\affiliation{Stanford Institute for Materials and Energy Sciences, SLAC National Accelerator Laboratory, Menlo Park, CA 94025, USA}
	\affiliation{Department of Physics, Stanford University, Stanford, CA 94305, USA}
	\affiliation{Department of Applied Physics, Stanford University, Stanford, CA 94305, USA}
	
	\author{Harold Y. Hwang}
	\affiliation{Stanford Institute for Materials and Energy Sciences, SLAC National Accelerator Laboratory, Menlo Park, CA 94025, USA}
	\affiliation{Department of Applied Physics, Stanford University, Stanford, CA 94305, USA}

	\author{Jigang Wang}\email{jgwang@ameslab.gov}
	\affiliation{Ames National Laboratory, Ames, IA 50011 USA}
	\affiliation{Department of Physics and Astronomy, Iowa State University, Ames, Iowa 50011, USA.}

\date{\today}

\begin{abstract}

\textbf{The dynamics of Higgs mode in superconductors, manifested as coherent oscillations of the superconducting order parameter amplitude, provides vital insights into the nature of the superconducting gap structure and symmetry. Here we utilize two-dimensional terahertz coherent spectroscopy to investigate Higgs dynamics of a newly discovered infinite-layer nickelate superconductor. While we observe distinct nonlinear terahertz responses from the superconducting state, well-defined long-lived Higgs modes, as commonly observed in $s$-wave superconductors, are entirely absent in the nickelate film. Instead, we find the coherent nonlinear terahertz response is dominated by the quasiparticle excitations. These observations strongly indicate that the Higgs mode in infinite-layer nickelates is heavily damped by the quasiparticle excitations at arbitrarily low energies, which is a characteristic of $d$-wave pairing symmetry. Additionally, by examining the temperature dependence of the nonlinear terahertz response, we discover short-range superconducting fluctuations in the vicinity of $T_\mathrm{c}$. Our findings provide proof of a new $d$-wave system and establish a foundation for investigating the unconventional superconductivity in nickelates.}  

\end{abstract}

\maketitle

\setlength{\parskip}{0.11em}

The quest for new superconductors that are structural and electronic analogs of cuprates is crucial to settle the long-standing puzzle of the pairing mechanism in high-$T_\mathrm{c}$ superconductors. The recent discovery of infinite-layer nickelate superconductors fills this void and expands the spectrum of unconventional superconductivity in oxides~\cite{nickelate_nature_2019}. Despite the exciting advances on the phase diagram and the charge/magnetic orders~\cite{nickelate_LAST_2022,nickelate_chargeorder,RXIS_magntic_2021,nickelate_chargespin_1,nickelate_chargespin_2}, to date, the \textit{bulk-sensitive} spectroscopic characterization associated with the superconducting transition in nickelates, in particular regarding the superconducting gap structure and symmetry, remains scarce. Due to their structural and electronic similarities with cuprates~\cite{nickelate_bandstr_Botana,nickelate_bandstr_Been}, it is plausible to expect a $d$-wave pairing symmetry in the nickelate superconductors~\cite{nickelate_dwaveprl_theory_2020,nickelate_dwaveprb_theory_2020,nickelate_pen_depth_Huang_2022,nickelate_dga_1,nickelate_dga_2,nickelate_dga_3}. In the meanwhile, a two-gap $s$-wave scenario has also been proposed, based on the multi-band character and the underestimated phonon--electron coupling strength in the nickelate superconductors~\cite{nickelate_theory_swave}. 

\begin{figure*}[t]

\includegraphics[clip,width=5.5in]{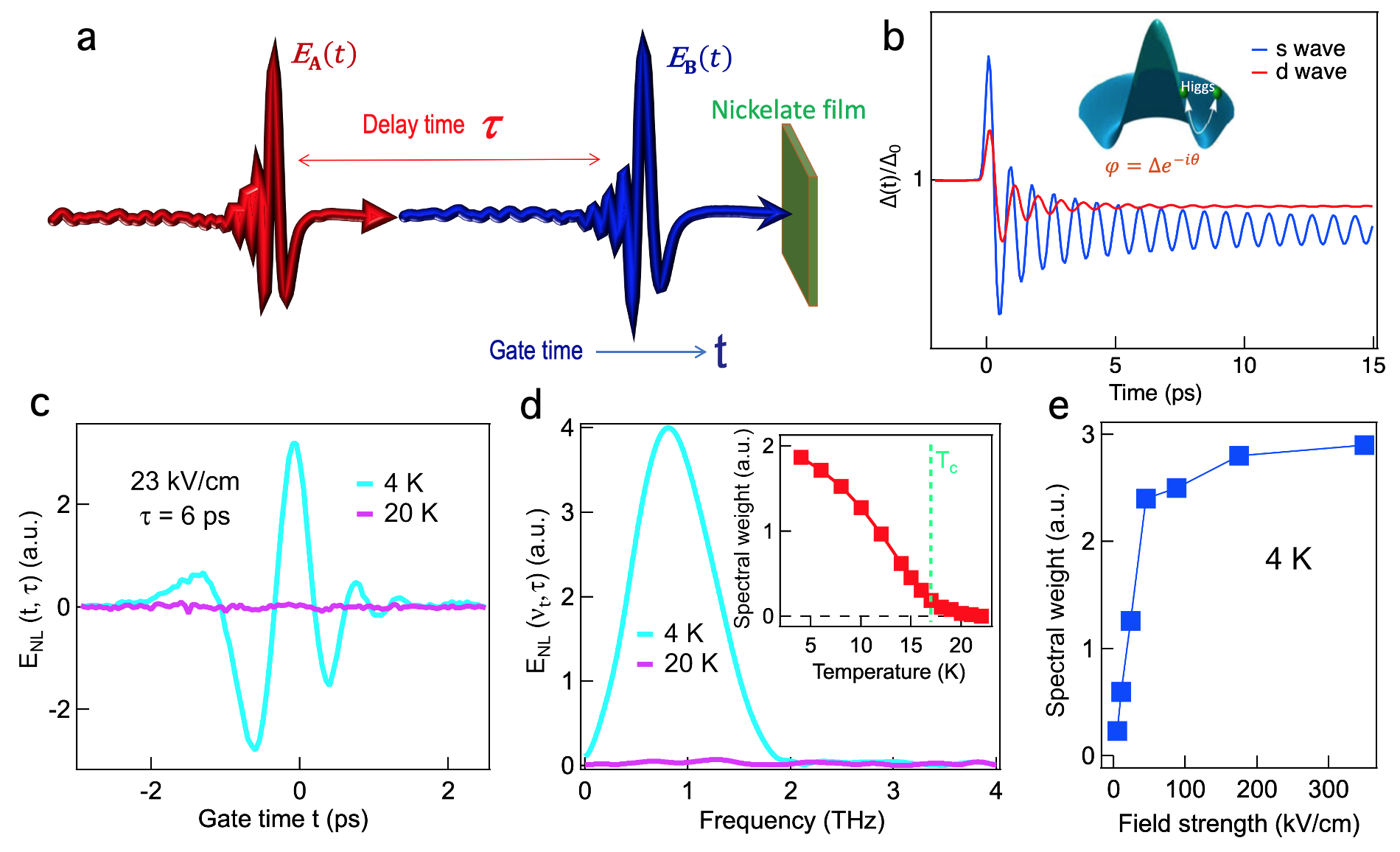}
\caption{ {\bf Two-dimensional THz coherent spectroscopic measurement of the nickelate superconducting film Nd$_{0.85}$Sr$_{0.15}$NiO$_2$.}  {\bf (a)} Experimental schematics. {\bf (b)} Comparison of the calculated non-equilibrium dynamics of the superconducting order parameter amplitude $\Delta(t)$/$\Delta_0$ between $s$-wave (blue line) and $d$-wave (red line) pairing symmetry. These amplitude fluctuations are induced by a single THz pump electric field in a clean-limit superconductor. The Higgs-mode oscillations are strongly damped in the $d$-wave system owing to the strong coupling between the amplitude fluctuations and the low energy quasiparticle excitations below 2$\Delta$. $\Delta_0$ is the superconducting gap amplitude in the equilibrium state. Inset: schematic illustration of the Higgs mode oscillations in the Mexican-hat-like free energy landscape. {\bf (c)} Temporal profiles of THz nonlinear signals $E_{\rm{NL}}(t,\tau)$ of the nickelate film at 4 and 20~K under the THz pump field strength of 23~kV/cm and the pulse separation of $\tau$ = 6~ps.  {\bf (d)} Corresponding FFT amplitude spectrum $E_{\rm{NL}}(\nu_t,\tau)$ at 4 and 20~K. The inset shows the integrated spectral weight $\int E_{\rm{NL}}(v,\tau)dv$ as a function of temperature which behaves like a superconducting order parameter. {\bf (e)} THz field strength dependence of the NL spectral weight $\int E_{\rm{NL}}(t,\tau)dt$ at 4 K.}

\label{Fig1}
\end{figure*}

The ambiguity in the nature of the superconducting gap structure and symmetry can be elucidated by investigating the Higgs modes and their dynamics which are inherently interconnected with the superconducting order parameter~\cite{Higgs_review1}. In $s$-wave superconductors, the Higgs mode in the long wavelength limit is usually a well-defined collective mode at the frequency of the superconducting gap 2$\Delta$~\cite{Higgs_review2,Higgs_2dTHz,THz_sc2}. As shown in Fig.~\ref{Fig1}b, the temporal $s$-wave Higgs oscillations in response to the THz pulse excitation of the superconducting order parameter are long-lived due to the absence of quasiparticle excitations below 2$\Delta$. This absence prevents the damping of the Higgs mode. In $d$-wave superconductors, however, the situation is completely different. Due to the lower rotational symmetry, the $d$-wave pairing symmetry in principle can enable Higgs modes with $A_{1g}$ and $B_{1g}$ symmetries~\cite{Higgs_1, Higgs_sdwave_theory1,Higgs_dwave_theory0}. The $A_{1g}$ Higgs mode of the $d$-wave order parameter, which oscillates at 2$\Delta$ in the long wavelength limit, is predicted to be detectable by using ultrafast spectroscopy~\cite{Higgs_dwave_thoery2, Higgs_1}.  Importantly, in $d$-wave superconductors the continuum of quasiparticle excitations extends to zero energy~\cite{dwave_arpes_1993,dwave_phasesensitive_1994}, which provides numerous rapid decay channels to significantly damp the Higgs excitations~\cite{Higgs_sdwave_theory1,Higgs_sdwave_arpes_theory2,Higgs_sdwave_theory3}. As a result, the $d$-wave Higgs modes will become extremely damped as compared with the $s$-wave Higgs modes, shown in our calculation in Fig.~\ref{Fig1}b. These contrasting behaviors in the temporal dynamics between $s$- and $d$-wave Higgs modes can serve as a unique probe of the superconducting gap structure and symmetry in nickelate superconductors.

\begin{figure*}[t]

\includegraphics[clip,width=5.5in]{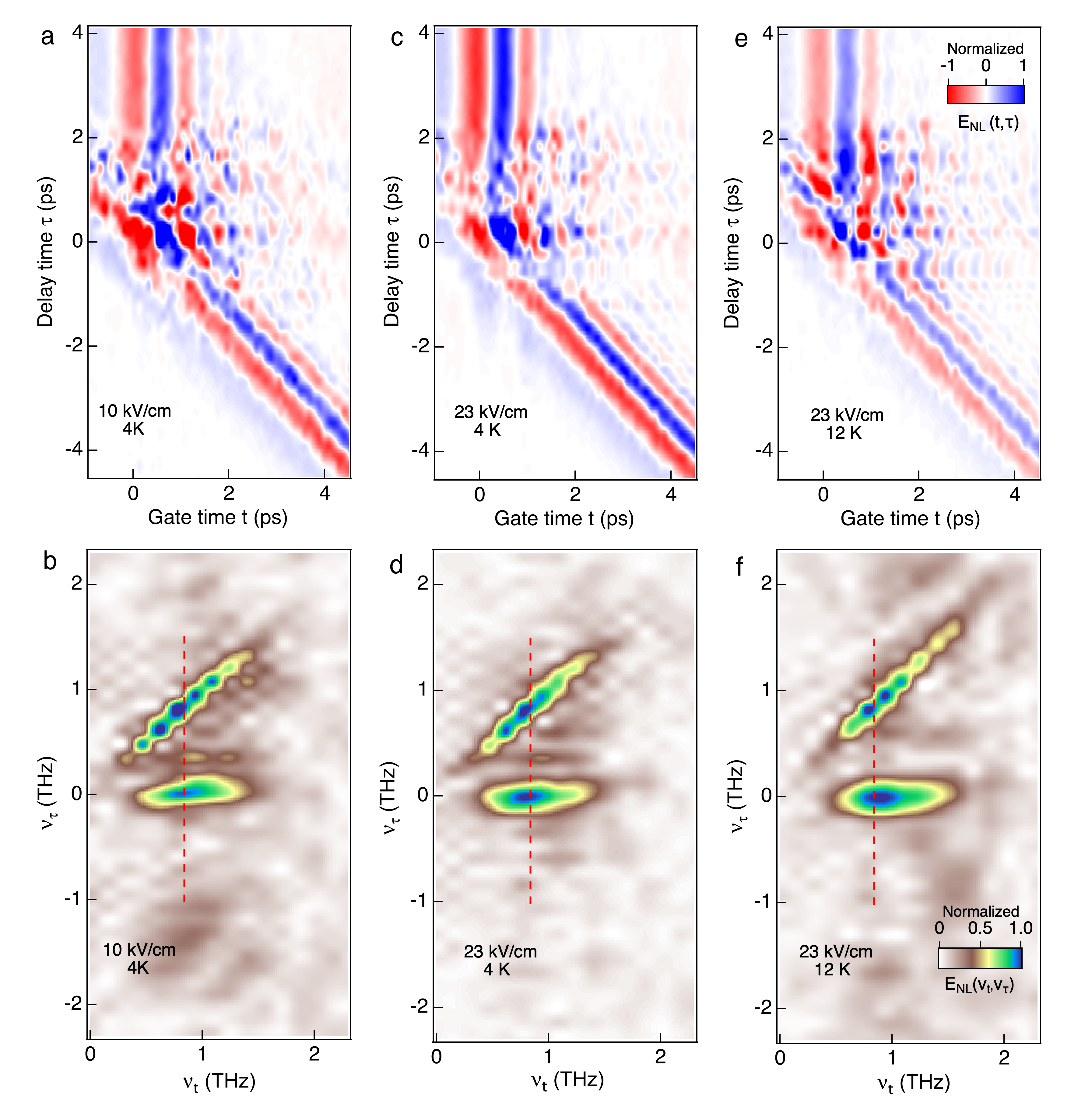}
\caption{ {\bf Two-dimensional THz coherent nonlinear spectra in superconducting state.} Two-dimensional false-color plot of the measured coherent nonlinear transmission $E_{\rm{NL}}(t,\tau)$ at {\bf (a)} 10 kV/cm, 4 K {\bf (c)} 23 kV/cm, 4 K, and {\bf (e)} 23 kV/cm, 12 ~K. Corresponding THz coherent spectra $E_{\rm{NL}}(\omega_t,\omega_{\tau}$) in 2D frequency space at {\bf (b)} 10 kV/cm, 4 K {\bf (d)} 23 kV/cm, 4 K, and {\bf (f)} 23 kV/cm, 12 K. The red dashed lines mark $\omega_{t}=0.84$ THz .}
\label{Fig2}
\end{figure*}

Here we employ two-dimensional (2D) terahertz (THz) coherent spectroscopy to investigate the superconducting pairing symmetry of an optimally doped nickelate film Nd$_{0.85}$Sr$_{0.15}$NiO$_2$ ($T_\mathrm{c}$ = 17~K)~\cite{nickelate_LAST_2022}. The nickelate films are grown on a THz transparent LSAT (001) substrate and exhibit excellent crystallinity and minimal defects~\cite{nickelate_LAST_2022}. Our results suggest that the Higgs mode in the nickelate film is highly damped owing to the presence of significant quasiparticle excitation continuum far below $2\Delta$, providing evidence for a $d$-wave pairing scenario.

2D THz coherent spectroscopy is a powerful spectroscopic tool to reveal the nature of collective excitations in quantum materials such as the Higgs mode in superconductors~\cite{Higgs_2dTHz} and the magnons in magnetic meterials~\cite{Magnon_2dTHz} by measuring their high-order nonlinear (NL) responses beyond conventional linear susceptibilities. As schematically illustrated in Fig.~\ref{Fig1}a, here two {\em collinear, phase-locked} THz pulses A and B with similar field strengths are focused on the nickelate film. Their temporal waveforms and the corresponding fast Fourier transform (FFT) spectra are presented in supplementary Fig. S1 (see Supplementary Note 1). The NL signal, $E_{\rm{NL}}(t,\tau)=E_{\rm{AB}}(t,\tau)-E_{\rm{A}}(t)-E_{\rm{B}}(t,\tau)$, is recorded as functions of both gate time $t$ and delay time $\tau$. Here $\tau$ is the time separation between pulses A and B. $E_{\rm{AB}}(t,\tau)$ is the transmitted signal when both pulse A and B are present. $E_{\rm{A}}(t)$ and $E_{\rm{B}}(t,\tau)$ are the transmitted signals with only pulse A or B present, respectively. The 2D FFT of the time-domain traces $E_{\rm{NL}}(t,\tau)$ with respect to $t$ and $\tau$ yields the amplitude spectrum $E_{\rm{NL}}(\nu_t,\nu_{\tau})$ in the 2D frequency space. The measured NL signal is generated by NL processes that involve excitations by both THz pulses, e.g., pump--probe, four-wave mixing, third harmonic generation, and higher order processes. These NL processes can be isolated in 2D frequency space via well-separated frequency vectors, e.g., $\textbf{$\nu$}_{\rm{A}} = (\nu, 0)$, and $\textbf{$\nu$}_{\rm{B}} = (\nu, \nu)$. Here $\nu$ could be the central frequency of the driving THz field, the Higgs collective mode frequency $2\Delta$, or the peak energy of the broad nonlinear quasiparticle excitations in superconductors.

\begin{figure*}[t]

\includegraphics[clip,width=5.8in]{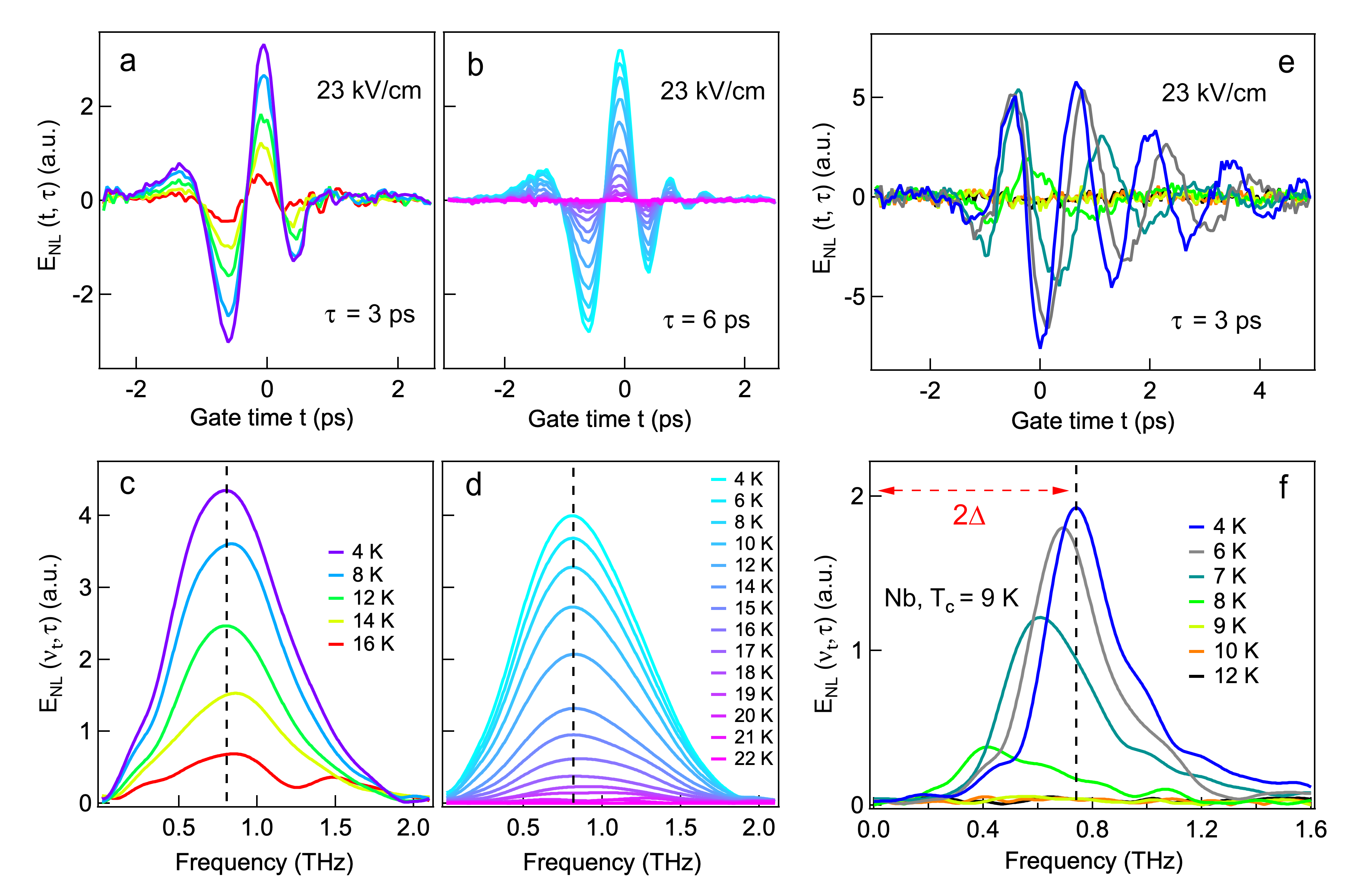}
\caption{ {\bf Temperature dependence of THz nonlinear signals at fixed pulse separations: comparison of nickelate and Nb s-wave superconductors}  Temporal profiles of the THz nonlinear signals $E_{\rm{NL}}(t,\tau)$ of the nickelate film at various temperatures under THz field strength of 23 kV/cm and pulse separation of {\bf (a)} $\tau=3$~ps and {\bf (b)}  $\tau = 6$~ps. Corresponding FFT amplitude spectrum $E_{\rm{NL}}(\nu_t,\tau)$ of the nickelate film at various temperatures under the THz field strength of 23 kV/cm and pulse separation of {\bf (c)} $\tau = 3$~ps and {\bf (d)}  $\tau= 6$~ps. {\bf (e)} Temporal profiles of THz nonlinear signals $E_{\rm{NL}}(t,\tau)$ in the case of a Nb film at various temperatures under THz field strength of 23 kV/cm and the pulse separation of $\tau = 3$~ps. {\bf (f)} Corresponding FFT amplitude spectrum $E_{\rm{NL}}(\nu_t,\tau)$ of the Nb film at various temperatures under THz pump field strength of 23~kV/cm and pulse separation of $\tau=3$~ps. Vertical dashed line indicates $2\Delta$ at $T=4$~K.}

\label{Fig3}
\end{figure*}

We present the measured NL signal $E_{\rm{NL}}(t,\tau)$ and the corresponding FFT amplitude spectrum $E_{\rm{NL}}(\nu,\tau)$ at 4~K and 20~K in Fig.~\ref{Fig1}c and \ref{Fig1}d. Here, the THz pulse time separation is fixed to $\tau = 6$~ps and both THz field strengths are $\sim 23$~kV/cm. The NL response to the two-pulse excitation in the superconducting state of the nickelate film manifests as a broad peak centered at $\sim$0.81 THz in the frequency domain. We calculate the integrated spectral weight $\int E_{\rm{NL}}(\nu,\tau)d \nu$ of NL response from data in Fig.~\ref{Fig3}d and present it as a function of temperature in the inset of Fig.~\ref{Fig1}d. The temperature dependence of this NL spectral weight gradually decreases to zero as the temperature approaches $T_\mathrm{c}$. Such an order-parameter-like behavior clearly demonstrates that the NL response below $T_\mathrm{c}$ is solely from nickelate superconductivity. In Fig.~\ref{Fig1}e, we show the THz field strength dependence of NL spectral weight at 4~K. The NL spectral weight first increases rapidly with THz field strength, and then saturates beyond 100~kV/cm. In order to study the Higgs dynamics of the nickelate film under the condition of partial THz quantum quench of the superconducting order parameter, we focus on the field strength below 50~kV/cm to avoid the saturation regime.

Figure~\ref{Fig2} presents the 2D temporal profile $E_{\rm{NL}}(t,\tau)$ and the corresponding 2D frequency spectra $E_{\rm{NL}}(\nu_t,\nu_{\tau})$ at different temperatures and THz field strengths. Here we emphasize three key points. First, at THz field strength of 10~kV/cm (Fig.~\ref{Fig2}b), the 2D frequency spectrum $E_{\rm{NL}}(\nu_t,\nu_{\tau})$ at 4~K shows two dominant peaks, at (0.84, 0) and at (0.84, 0.84), along the $\nu_{t}=0.84$-axis marked by the red dashed line. These two peaks are generated by the third-order NL process of AB pump probe, ${\nu}_{\rm{A}}-{\nu}_{\rm{A}}+{\nu}_{\rm{B}}$, and BA pump probe, ${\nu}_{\rm{B}}-{\nu}_{\rm{B}}+{\nu}_{\rm{A}}$~\cite{Higgs_2DTHz_theory}, respectively. We do not observe NL responses from four-wave mixing processes such as 2${\nu}_{\rm{A}}-{\nu}_{\rm{B}}$ and $2{\nu}_{\rm{B}}-{\nu}_{\rm{A}}$~\cite{Higgs_2DTHz_theory}. Second, to investigate if the observed pump--probe peaks arise from Higgs mode excitations, we compare to the peaks obtained for a higher  THz field strength of 23 kV/cm. As demonstrated by our previous 2D THz study of an iron-based superconducting film, the higher field strength will partially quench the superconducting gap 2$\Delta$, which yields a redshift of the Higgs mode frequency~\cite{Higgs_2dTHz}. As shown in Fig.~\ref{Fig2}d, we do not observe any clear shifts of the pump--probe peaks in the 2D frequency space with increasing field, which indicates that the peak positions are not determined by the Higgs mode frequency 2$\Delta$. Third, we keep the field strength at 23 kV/cm and raise the measurement temperature to 12 K. With increasing temperature, the superconducting gap shrinks, which will result in a redshift of the Higgs mode frequency too\cite{Higgs_2dTHz}. However, as shown in Fig.~\ref{Fig2}f, when compared to the 2D spectra at 4 K, the two pump--probe peaks at 12 K do not show shifts either, another indication that they do not follow the temperature dependence of the Higgs mode frequency 2$\Delta$. These observations suggest that the pump--probe peaks are not generated by the Higgs mode excitations. The sharp Higgs response is completely absent in the nickelate superconductor. Instead, the broad quasiparticle excitations dominate the NL THz responses.

\begin{figure*}[t]

\includegraphics[scale=0.43]{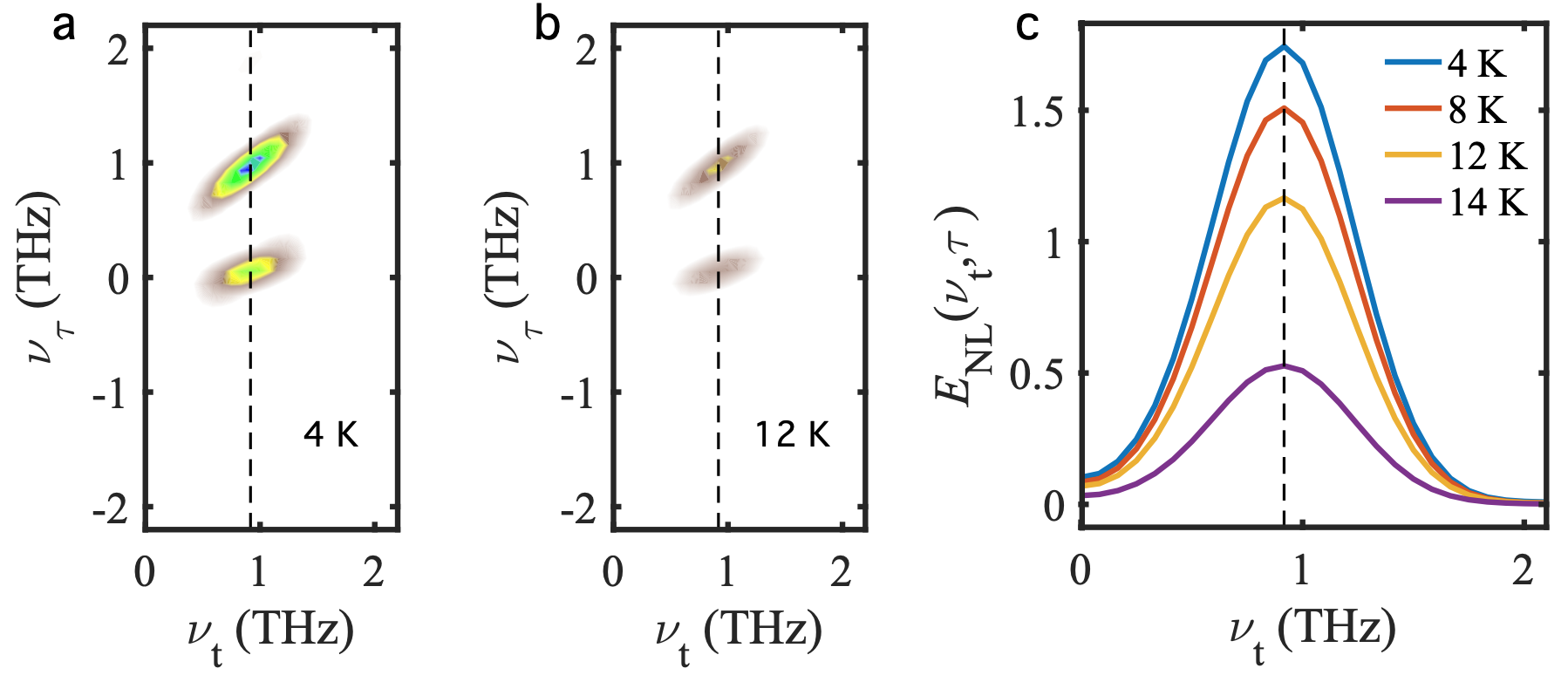}
\caption{ {\bf Calculated 2D THz spectra of $d$-wave superconductors.}  Two-dimensional false-color plot of
	the theoretical nonlinear coherent transmission spectrum $E_\mathrm{NL}(\nu_t,\nu_\tau)$ for a $d$-wave superconductor at temperatures of ($\textbf{a}$) 4~K and ($\textbf{b}$) 12~K. The vertical dashed lines indicate the position of the dominant pump--probe signals. ($\textbf{c}$) Calculated $E_\mathrm{NL}(\nu_t,\tau)$ spectra for various temperatures at fixed pulse separation $\tau=3$~ps. In the case of $d$-wave order parameter symmetry, the peak position (dashed line) is insensitive to the temperature, unlike the case of $s$-wave symmetry \cite{Mootz2022}}

\label{Fig4}
\end{figure*}

%which is expected in $d$-wave superconductors where the Higss mode strongly couples with a continuum of quasiparticle excitations. 

Our measured 2D THz spectra of nickelate superconductor imposes strong constraints on the possible superconducting gap symmetries. In $s$-wave superconductors, no matter one-band \cite{Higgs_2DTHz_theory} or multi-band \cite{Higgs_2dTHz}, their Higgs modes appear as well-defined collective modes robust to impurity scattering~\cite{swave_Higgs_dirty1,swave_Higgs_dirty2, laura}. The $s$-wave Higgs mode behavior has been extensively demonstrated in a number of systems such as Ba(Fe$_{1-x}$Co$_{x}$)$_2$As$_2$~\cite{THz_sc2}, NbN~\cite{NbN_Higgs}, and Nb$_{1-x}$Ti$_{x}$N~\cite{NbTiN_Higgs}. It is worth noting that a recent 2D THz measurement of Ba(Fe$_{0.92}$Co$_{0.08}$)$_2$As$_2$, a multi-band $s$-wave superconductor, has revealed rich Higgs mode dynamics that changes drastically as functions of THz driving field strength and temperature~\cite{Higgs_2dTHz,Higgs_2DTHz_theory}. The $s$-wave Higgs mode is found to participate in multiple high-order NL processes, resulting in clear Higgs sidebands in $E_{\rm{NL}}(\nu_t,\nu_{\tau})$. The absence of sharp Higgs dynamics in the 2D THz spectra of the nickelate superconductor is inconsistent with an $s$-wave scenario. Instead, it strongly suggests the existence of quasiparticle excitations at arbitrarily low energies, which significantly damp Higgs mode and suppress its contribution to the 2D THz response. The observation of highly damped Higgs mode behaviors provides compelling evidence in favor of a $d$-wave scenario in the nickelate superconductor. Actually, in cuprates, despite the theoretical predictions of multiple Higgs modes, ultrafast measurements using THz pump-probe spectroscopy and time-resolved photoemission spectroscopy have never observed clear \textit{free} coherent Higgs oscillations due to the strong damping from the low-energy quasiparticle excitations. Recent 2D THz measurements of La$_{2-x}$Sr$_x$CuO$_4$ and Bi$_2$Sr$_2$CaCu$_2$O$_{8+x}$ films display very similar NL THz behaviors as in the nickelates~\cite{Dipanjan}, e.g., the absence of sharp Higgs dynamics, the dominance of the quasiparticle pump-probe response below $T_\mathrm{c}$, and the absence of the temperature-driven redshift of NL peaks. These strong similarities between nickelate superconductors and cuprates further support the $d$-wave scenario in the current nickelate film.

We now present additional experimental evidence to support the claim that the Higgs mode in the nickelate film is highly damped. In Fig.~\ref{Fig3}a to \ref{Fig3}d, we show the temperature dependent NL signal $E_{\rm{NL}}(t,\tau)$ and the corresponding $E_{\rm{NL}}(\nu_t,\tau)$ at two fixed pulse time separations, $\tau = 3$ and $6$ ps. As discussed earlier, the Higgs mode frequency will redshift with increasing temperature. If any long—lived Higgs modes exist, the spectral weight of $E_{\rm{NL}}(\nu_t,\tau)$ should gradually move to lower frequency with increasing temperature. At the field strength of 23 kV/cm, as shown in Fig.~\ref{Fig3}c and \ref{Fig3}d, we do not observe any spectral weight transfer between different frequencies in  $E_{\rm{NL}}(\nu_t,\tau)$ except the uniform decrease of the NL signal with increasing temperature. The peak position of $E_{\rm{NL}}(\nu_t,\tau)$ (dashed black line) does not move at all. More temperature dependent data at field strengths of 5, 10, and 45 kV/cm are presented in supplementary Fig. S5 and S6 (see Supplementary Note 5), which demonstrate similar temperature behaviors at 23 kV/cm. The lack of sharp Higgs dynamics and the temperature-induced spectral weight shifts of $E_{\rm{NL}}(\nu_t,\tau)$ here is fully consistent with the $d$-wave scenario, which has been clearly demonstrated in cuprates by the 2D THz measurements~\cite{Dipanjan}.

To make our comparison more clearly, in Fig. \ref{Fig3}e and \ref{Fig3}f, we present the temperature-dependent NL signal $E_{\rm{NL}}(t,\tau)$ and $E_{\rm{NL}}(\nu_t,\tau)$ of an $s$-wave element superconductor Nb ($T_\mathrm{c}$ = 9 K) at $\tau = 3$ ps under the same field strength. At 4 K, $E_{\rm{NL}}(\nu_t,\tau)$ shows a sharp Higgs resonance peak at 0.74 THz which coincides with the superconducting gap 2$\Delta$~\cite{Nb_SC_gap}. With increasing temperature, the Higgs resonance peak redshifts as $2 \Delta$, and ultimately disappears above $T_\mathrm{c}$. Within the entire measured temperature range, the spectral weight of $E_{\rm{NL}}(\nu_t,\tau)$ shows a significant temperature-driven shift between different frequency sectors, which is in sharp contrast to the  behaviors observed in the nickelate superconductor.

To further support the $d$-wave scenario in the nickelate superconductor, we simulate the 2D THz coherent spectra of a $d$-wave system by solving the gauge-invariant superconducting Bloch equations including the electron--impurity scattering~\cite{Mootz2020,Higgs_2DTHz_theory}. We also take into account the electromagnetic propagation effects in a superconductor which dynamically generates a DC supercurrent~\cite{vaswani2019discovery}. This leads to distinct Higgs mode peaks in the 2D THz spectra which have been experimentally observed in Ba(Fe$_{1-x}$Co$_{x}$)$_2$As$_2$~\cite{Higgs_2dTHz}. The details of our theory can be found in Supplementary Note 7 to 9. Previous magneto transport measurements of nickelate films reveal that the nickelate superconductivity falls in the dirty limit\cite{nickelate_dirtylimit}. Here, to qualitatively capture the features of our experimental data, we choose the impurity scattering rate $\Gamma$ = 3 THz and the $d$-wave gap amplitude 2$\Delta$ = 1.2 THz in our numerical simulations to fulfill the dirty-limit condition. Figure~\ref{Fig4}a and \ref{Fig4}b present the simulated nonlinear 2D THz spectra, $E_\mathrm{NL}(\nu_t,\nu_\tau)$, for temperatures of 4 and 12~K, respectively.  The 2D THz spectra show two strong pump--probe peaks approximately located at $(0.9,0.9)$ and $(0.9,0.0)$ (vertical dashed lines), qualitatively in agreement with the experimental results in Fig.~\ref{Fig2}. These calculated 2D THz signals result from the quasiparticle excitations and decrease in strength with growing temperature, while their positions remain unchanged~\cite{Higgs_2DTHz_theory}. In contrast to our 2D THz simulations of $s$-wave superconductors~\cite{Higgs_2DTHz_theory,Higgs_2dTHz}, the 2D spectra in Fig.~\ref{Fig4}a and \ref{Fig4}b do not show any Higgs mode signals at the Higgs mode frequency 2$\Delta$. The absence of such signals is further confirmed by Fig.~\ref{Fig4}c, where the $E_\mathrm{NL}(\nu_t,\tau)$ spectra are shown for various temperatures at fixed $\tau=3$~ps. With increasing temperature, the peak position of the calculated $E_\mathrm{NL}(\nu_t,\tau)$ spectra does not move and no spectral weight shifts are observable, in agreement with the experimental results in Fig.~\ref{Fig3}c and \ref{Fig3}d. The agreement of these theoretical and experimental findings provides additional evidence for the $d$-wave scenario in nickelates.

\begin{figure}[t]

\includegraphics[clip,width=3.2in]{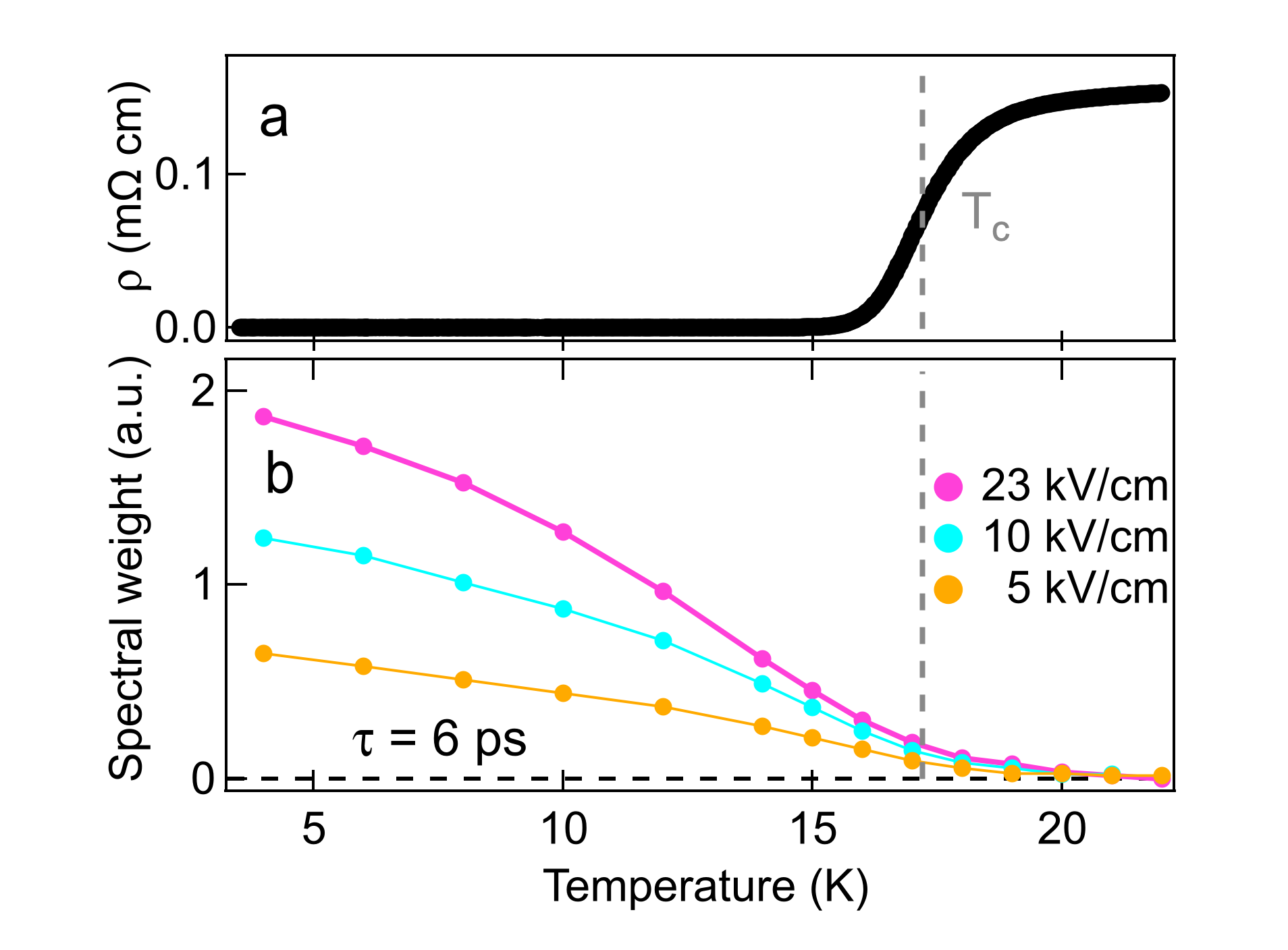}
\caption{ {\bf Superconducting fluctuations revealed by THz nonlinear response.} ($\textbf{a}$) DC resistivity of the nickelate film. ($\textbf{b}$) The integrated spectral weight $\int E_{\rm{NL}}(\nu,\tau)d \nu$ of the 2D THz nonlinear response as a function of temperature under THz field strengths of 5, 10 and 23 kV/cm, and pulse time separation of $\tau = 6$ ps.}
\label{Fig5}
\end{figure}

Finally, we discuss the role of superconducting fluctuations in the nickelate film. As shown in Fig.~\ref{Fig3}, and supplementary Fig. S5 and S6 (see Supplementary Note 5), the NL THz response persists beyond $T_\mathrm{c}$ $\sim$ 17 K. To quantitatively track this temperature dependence, we show the integrated spectral weight $\int E_{\rm{NL}}(\nu,\tau)dv$ as a function of temperature at three THz pump field strengths of 5, 10, and 23 kV/cm in Fig.~\ref{Fig5}b. The DC resistivity $\rho$ is plotted in Fig.~\ref{Fig5}a for comparison. Below $T_\mathrm{c}$, the NL THz responses exhibit a superconducting order-parameter-like behavior. Across $T_\mathrm{c}$, however, the NL signal still shows temperature and THz field strength dependence up to 20 K. A recent London penetration depth measurement using kHz inductance coils showed that the long-range coherence of Cooper pairs vanishes at the temperature where $\rho$ completely drops to zero~\cite{nickelate_pen_depth_Huang_2022}. The low-frequency measurements such as the kHz penetration depth measurement could track the macroscopic superconducting coherence. However, they cannot capture the rapid superconducting fluctuations on a timescale of picoseconds. The NL THz measurement has been demonstrated to be more sensitive to such rapid temporal superconducting correlation\cite{NbN_2dTHz_2022}. The NL THz response observed by 2D THz coherent spectroscopy near $T_c$ could be attributed to the short-range superconducting fluctuations in the nickelate film.

In conclusion, we use 2D THz coherent spectroscopy to study the Higgs dynamics in a newly discovered nickelate superconductor. Neither the NL THz response in the 2D frequency space nor the detailed temperature dependent NL signals resolve well-defined Higgs mode signals. These results demonstrate the existence of a continuum of quasiparticle excitations at arbitrarily low energies that couples strongly to the Higgs mode, resulting in highly damped Higgs mode behaviors in the infinite-layer nickelates. Our observation is consistent with the recent theoretical prediction of a $d_{x^2-y^2}$-wave pairing symmetry in nickelate superconductors~\cite{nickelate_dwaveprb_theory_2020,nickelate_dwaveprl_theory_2020}. We further discuss the superconducting fluctuations in the nickelate film.

\footnotesize

\bigskip

\section{Methods}

\textbf{2D THz coherent spectroscopy.}  A home-built 2D THz coherent spectrometer driven by a 1 kHz Ti : sapphire regenerative amplifier, which has 800 nm central wavelength and 40 fs pulse duration, is used for this work. The laser is split into three beams. Two are used to generate a pair of intense phase-locked THz pulses up to 350 kV/cm that are separated by a time delay $\tau$ using the tilted pulse front technique. The third beam is used to detect phase-locked THz electric fields in time-domain via standard electro-optic sampling. The two intense THz pulses are chopped with a frequency of 250 and 500 Hz respectively, which enables us to isolate the nonlinear correlated signal $E_{\rm{NL}}(t,\tau)=E_{\rm{AB}}(t,\tau)-E_{\rm{A}}(t)-E_{\rm{B}}(t,\tau)$. Here, $\tau$ is the time separation between pulse A and B. $t$ is the gate time of electro-optic sampling. $E_{\rm{AB}}(t,\tau)$ is the transmitted signal when pulse A and B are both present. $E_{\rm{A}}(t)$ and $E_{\rm{B}}(t,\tau)$ are the transmitted signals with only pulse A or B present, respectively. A few THz polarizers are placed inside the spectrometer to control the THz electric field strength.  

\textbf{Sample growth.}  $\sim$15 unit cells of perovskite Nd$_{0.85}$Sr$_{0.15}$NiO$_2$ thin films ($\sim$ 4.3 nm) were synthesized by pulsed-laser deposition with a KrF excimer laser ($\lambda$ = 248 nm) on 5 $\times$ 5 mm$^2$ LSAT (001) substrates, with the growth conditions specified in Ref. \cite{nickelate_LAST_2022}. The film was capped by a $\sim$ 4 unite cells of SrTiO$_3$ layer. The samples were then cut into two 2.5 $\times$ 5 mm$^2$ pieces and vacuum-sealed ($<$ 0.1 mTorr) in a Pyrex glass tube with $\sim$0.1 g of CaH$_2$ powder after loosely wrapping with aluminum foil. The glass tube was heated at 260 $^{\circ}$C for $\sim$2.5 hours, with temperature ramp rate of 10 $^{\circ}$C min$^{-1}$, to achieve topotactic transition to infinite-layer Nd$_{0.85}$Sr$_{0.15}$NiO$_2$~\cite{nickelate_nature_2019,nickelate_LAST_2022}. Temperature-dependent resistivity measurements were done in a six-point Hall bar geometry using aluminum wire-bonded contacts.

\textbf{Model Hamiltonian}

In this paper, we use a microscopic model of electron--impurity scattering~\cite{Higgs_dwave_thoery2}. We start from the real space Hamiltonian
\begin{align}
\label{eq:Ham}
&H=H_\mathrm{BdG}+H_\mathrm{e-i}\,,\nonumber \\
&H_\mathrm{BdG}=\sum_{\alpha}\int\mathrm{d}^3\mathbf{x}\,\psi_{\alpha}^\dagger(\mathbf{x})\left[\varepsilon(\mathbf{p}+e\mathbf{A}(\mathbf{x},t))-\mu-e\phi(\mathbf{x},t) \right. \nonumber \\
&\left.\qquad\qquad\qquad\qquad\qquad\qquad+\mu_\mathrm{H}(\mathbf{x})+\mu^{\alpha}_\mathrm{F}(\mathbf{x})\right]\psi_{\alpha}(\mathbf{x}) \nonumber \\
&\qquad\qquad -\int\mathrm{d}^3\mathbf{x}\left[\Delta(\mathbf{x})\psi^\dagger_{\uparrow}(\mathbf{x})\psi^\dagger_{\downarrow}(\mathbf{x})+\mathrm{h.c.}\right]\,,\nonumber \\
&H_\mathrm{e-i}=n_\mathrm{i}\sum_{\alpha}\int\mathrm{d}^3\mathbf{x}\,\psi_{\alpha}^\dagger(\mathbf{x})V(\mathbf{x})\psi_{\alpha}(\mathbf{x})\,.
\end{align}
Here, $\psi_\alpha^\dagger(\mathbf{x})$ and $\psi_\alpha(\mathbf{x})$ represent the electron creation and annihilation operators, respectively, with $\alpha$ denoting the spin index. The real space Bogolubov--de Gennes Hamiltonian, $H_\mathrm{BdG}$, contains the complex superconducting order parameter $\Delta(\mathbf{x})=-2\,g\langle \psi_{\downarrow}(\mathbf{x})\psi_{\uparrow}(\mathbf{x})\rangle=|\Delta(\mathbf{x})|\mathrm{e}^{\mathrm{i}\theta(\mathbf{x})}$, where $g$ represents the effective electron--electron pairing interaction and $\theta(\mathbf{x})$ denotes the phase of the spatially—dependent order parameter. The energy band dispersion is $\varepsilon(\mathbf{p})$, where $\mathbf{p}=-\mathrm{i}\nabla_\mathbf{x}$ ($\hbar=1$) is the momentum operator, $-e$ is the electron charge, and $\mu$ represents the equilibrium chemical potential. The coupling of the SC system to the electromagnetic fields is described by the vector potential $\mathbf{A}(\mathbf{x},t)$  and the scalar potential $\phi(\mathbf{x},t)$. The Fock energy 
%---------------------------
\begin{align}
\mu^{\alpha}_\mathrm{F}(\mathbf{x})=-g n_{\alpha}(\mathbf{x})\,,
\label{eq:mu_F}
\end{align}
%---------------------------	
Ensures charge conservation of the SC systems, whereas the Hartree energy,
\begin{align}
&\mu_\mathrm{H}(\mathbf{x})=2\sum_\sigma\int\mathrm{d}^3\mathbf{x}'\,V(\mathbf{x}-\mathbf{x}')n_{\sigma}(\mathbf{x}')\,,\nonumber \\
&n_{\sigma}(\mathbf{x})=\langle\psi^\dagger_{\sigma}(\mathbf{x})\psi_{\sigma}(\mathbf{x})\rangle\,,
\label{eq:mu_H}
\end{align}
moves the in-gap phase mode of the SC order parameter up to the plasma frequency, which is located inside the quasiparticle continuum. Here, $V(\mathbf{x})$ is the Coulomb potential whose Fourier transformation is given by $V_\mathbf{q}=e^2/(\varepsilon_0 q)$. The second part of the Hamiltonian~(\ref{eq:Ham}), $H_\mathrm{e-i}$,  describes electron--impurity scattering with impurity density $n_\mathrm{i}$, which determines the strength of the electron--impurity scattering.

%\section{Quantum kinetic equations}
\textbf{ Gauge Invariant Quantum kinetic equations}

To investigate the non-equilibrium dynamics of superconductors under THz excitation, we use the gauge-invariant density matrix approach presented in Refs.~\cite{Mootz2020,Higgs_2DTHz_theory}. As discussed at length in our previous publications, the advantage of this method as compared to other approaches is that it allows us to treat non-perturbatively the effects of amplitude and phase order parameter fluctuations, charge fluctuations, electromagnetic propagation effects, and spatial fluctuations. All of the above effects have been found to be importnat for explaining various aspects of the experimental observations demonstrating the break down of susceptibility perturbative expansions,  the dynamical breaking of inversion symmetry, and light—induced phase—amplitude nonlinear coupling. The obtained gauge-invariant Bloch equations for the Wigner function $\rho(\mathbf{k},\mathbf{R})$ depend on both momentum $\mathbf{k}$ and center-of-mass coordinate $\mathbf{R}$, where the latter describes the spatial fluctuations. In this paper, we assume weakly spatially-dependent superconducting systems where we can disregard all orders $\mathcal{O}(\nabla_\mathbf{k}\cdot\nabla_\mathbf{R})$ in the gradient expansion of the full spatially-dependent SC Bloch equations~\cite{Mootz2020}. We also assume homogeneous $E$-fields by neglecting their $\mathbf{R}$-dependence. Furthermore, since the Hartree potential contribution to the nonlinear response is small for weak spatial dependence, we set $\mu_\mathrm{H}=0$. The microscopic electron--impurity scattering described by $H_\mathrm{e-i}$ is included on the level of second-Born and Markov approximations in the density matrix equations of motion~\cite{Higgs_dwave_thoery2,PQE}, which allows for a quantum kinetic treatment of the disorder effects. 
To clarify the non-equilibrium dynamics of superconductors, we employ the pseudo-spin formalism introduced by Anderson in Ref.~\cite{Anderson}. In this formalism, the gauge-invariant density matrix is expressed in terms of Anderson pseudo-spin components at each wavevector $\mathbf{k}$:
%---------------------------------------
\begin{align}
\label{eq:ps}
\rho(\mathbf{k})=\sum_{n=0}^3 \rho_n(\mathbf{k})\sigma_n\,,
\end{align}
%---------------------------------------
where $\sigma_n$ ($n=1 \cdots 3$) are the Pauli spin matrices, $\sigma_0$ is the unit matrix, and $\rho_n(\mathbf{k})$ are the pseudo-spin components. For a weakly spatially-dependent superconducting system, at the lowest order in the gradient expansion of spatial fluctuations,  the gauge-invariant Bloch equations expressed in terms of pseudo-spins are given by~\cite{Mootz2020,Higgs_2DTHz_theory}:
%---------------------------------------
\begin{align}
\label{eq:eom_ps}
\frac{\mathrm{d}}{\mathrm{d}t}\rho_0(\mathbf{k})&=-e\mathbf{E}\cdot\nabla_\mathbf{k}\rho_3(\mathbf{k})+|\Delta_\mathbf{k}|\left[\rho_2(\mathbf{k}+\mathbf{p}_\mathrm{S}/2)-\rho_2(\mathbf{k}-\mathbf{p}_\mathrm{S}/2)\right]\nonumber \\
&+\frac{\mathrm{d}}{\mathrm{d}t}\rho_0(\mathbf{k})\bigg|_\mathrm{e-i}\,, \nonumber \\
%----------------
\frac{\mathrm{d}}{\mathrm{d}t}\rho_1(\mathbf{k})&=-\left[\varepsilon(\mathbf{k}-\mathbf{p}_\mathrm{S}/2)+\varepsilon(\mathbf{k}+\mathbf{p}_\mathrm{S}/2)+2\,\mu_\mathrm{eff}(t)+2\mu_\mathrm{F}(t)\right]\rho_2(\mathbf{k}) \nonumber \\
&+\frac{\mathrm{d}}{\mathrm{d}t}\rho_1(\mathbf{k})\bigg|_\mathrm{e-i}\,, \nonumber \\
%----------------
\frac{\mathrm{d}}{\mathrm{d}t}\rho_2(\mathbf{k})&=\left[\varepsilon(\mathbf{k}-\mathbf{p}_\mathrm{S}/2)+\varepsilon(\mathbf{k}+\mathbf{p}_\mathrm{S}/2)+2\,\mu_\mathrm{eff}(t)+2\mu_\mathrm{F}(t)\right]\rho_1(\mathbf{k}) \nonumber \\
&+|\Delta_\mathbf{k}|\left[\rho_3(\mathbf{k}+\mathbf{p}_\mathrm{S}/2)+\rho_3(\mathbf{k}-\mathbf{p}_\mathrm{S}/2)\right.\nonumber \\
&\left.\qquad\quad-\rho_0(\mathbf{k}-\mathbf{p}_\mathrm{S}/2)+\rho_0(\mathbf{k}+\mathbf{p}_\mathrm{S}/2)\right]\nonumber \\ &+\frac{\mathrm{d}}{\mathrm{d}t}\rho_2(\mathbf{k})\bigg|_\mathrm{e-i}\,, \nonumber \\
%----------------
\frac{\mathrm{d}}{\mathrm{d}t}\rho_3(\mathbf{k})&=-e\mathbf{E}\cdot\nabla_\mathbf{k}\rho_0(\mathbf{k})-|\Delta_\mathbf{k}|\left[\rho_2(\mathbf{k}+\mathbf{p}_\mathrm{S}/2)+\rho_2(\mathbf{k}-\mathbf{p}_\mathrm{S}/2)\right]\nonumber \\ &+\frac{\mathrm{d}}{\mathrm{d}t}\rho_3(\mathbf{k})\bigg|_\mathrm{e-i}\,.
\end{align}
In the above equations, the effective driving field ${\bf E}(t)$ differs from the laser field due to electromagnetic propagation effects as determined by Maxwell’s equations. Importantly, the above equations describe non—perturbatively the dynamical coupling between charge fluctuations ($\rho_0$ and pseudo—spin fluctuations ($\rho_3$ induced by the electromagnetic field, which leads to condensate quantum transport effects that arise to linear order in the driving electric field (paramagnetic current contribution)and have been discussed at length in our previous publications. 
For a  homogeneous system, the time-dependent SC order parameter amplitude is given by 
\begin{align}
|\Delta_\mathbf{k}|=-2\sum_{\mathbf{k}'}g_{\mathbf{k},\mathbf{k}'}\rho_1(\mathbf{k}')\,.
\label{gap} 
\end{align} 
The effective time—dependent chemical potential $\mu_\mathrm{eff}(t)$ is determined by the time-varying phase $\theta(t)$ of the SC order parameter and by the scalar potential $\phi(t)$ induced by external electromagnetic fields according to
\begin{align} 
\label{eff-pot} 
\mu_\mathrm{eff}(t)=e\,\phi(t)+\frac{1}{2}\frac{\partial}{\partial t}\theta(t)-\mu\,.
\end{align} 
The acceleration of Cooper pairs by the electric field results in a superfluid momentum $\mathbf{p}_\mathrm{S}(t)=-2\,e\,\mathbf{A}(t)$, which is modified by spatial fluctuations in the effective chemical potential.

In contrast to conventional pseudo-spin models~\cite{Higgs_swave_theory2,Aoki2017,Forster2017,Higgs_sdwave_theory1}, Eq.(\ref{eq:eom_ps}) contains transport terms like $e\,\mathbf{E}\cdot\nabla_\mathbf{k}\rho_3(\mathbf{k})$, which cause displacement of electronic populations in ${\bf k}$-space due to condensate acceleration.
The coupling between $\rho_0(\mathbf{k})$ and $\rho_3(\mathbf{k})$ in Eq.~(\ref{eq:eom_ps}) arises from inversion symmetry breaking induced by the THz field ${\bf E}(t)$. 
The latter can result in symmetry-forbidden harmonics, gapless superconductivity, and quasi-particle quantum phases controlled by strong THz field pulses~\cite{THz_sc3,yang2019lightwave,vaswani2019discovery,Mootz2020}.

The electron--impurity scattering contributions to Eq.~(\ref{eq:eom_ps}) are given by ~\cite{Higgs_dwave_thoery2}
\begin{align}
\label{eq:scat}
&\frac{\mathrm{d}}{\mathrm{d}t}\rho_0(\mathbf{k})\bigg|_\mathrm{e-i} \nonumber \\
&=\pi n_\mathrm{i}\sum_{\mathbf{k}'}V^2_{\mathbf{k}-\mathbf{k}'}\left[\rho_0(\mathbf{k}')-\rho_0(\mathbf{k})\right]\left[\delta(E_{\mathbf{k}}-E_{\mathbf{k}'})+\delta(E_{\mathbf{k}}+E_{\mathbf{k}'})\right. \nonumber \\ &\qquad\qquad\;\left. +\left(\delta(E_{\mathbf{k}}-E_{\mathbf{k}'})-\delta(E_{\mathbf{k}}+E_{\mathbf{k}'})\right)\left(l^2_{\mathbf{k},\mathbf{k}'}-m^2_{\mathbf{k},\mathbf{k}'}\right)\right] \nonumber \\
&\frac{\mathrm{d}}{\mathrm{d}t}\rho_1(\mathbf{k})\bigg|_\mathrm{e-i} \nonumber \\
&=-\pi n_\mathrm{i}\sum_{\mathbf{k}'}V^2_{\mathbf{k}-\mathbf{k}'}\{\left[\rho_1(\mathbf{k}')+\rho_1(\mathbf{k})\right]\left[\delta(E_{\mathbf{k}}-E_{\mathbf{k}'})+\delta(E_{\mathbf{k}}+E_{\mathbf{k}'})\right. \nonumber \\ &\qquad\qquad\;\left. +\left(\delta(E_{\mathbf{k}}-E_{\mathbf{k}'})-\delta(E_{\mathbf{k}}+E_{\mathbf{k}'})\right)\left(l^2_{\mathbf{k},\mathbf{k}'}-m^2_{\mathbf{k},\mathbf{k}'}\right)\right] \nonumber \\
&+2\left[\rho_3(\mathbf{k}')+\rho_3(\mathbf{k})\right]\left(\delta(E_{\mathbf{k}}-E_{\mathbf{k}'})-\delta(E_{\mathbf{k}}+E_{\mathbf{k}'})\right)m_{\mathbf{k},\mathbf{k}'}l_{\mathbf{k},\mathbf{k}'}\} \nonumber \\
&\frac{\mathrm{d}}{\mathrm{d}t}\rho_2(\mathbf{k})\bigg|_\mathrm{e-i}\nonumber \\
&=-\pi n_\mathrm{i}\sum_{\mathbf{k}'}V^2_{\mathbf{k}-\mathbf{k}'}\left[\rho_2(\mathbf{k}')+\rho_2(\mathbf{k})\right]\left[\delta(E_{\mathbf{k}}-E_{\mathbf{k}'})+\delta(E_{\mathbf{k}}+E_{\mathbf{k}'})\right. \nonumber \\ &\qquad\qquad\left. +\left(\delta(E_{\mathbf{k}}-E_{\mathbf{k}'})-\delta(E_{\mathbf{k}}+E_{\mathbf{k}'})\right)\left(l^2_{\mathbf{k},\mathbf{k}'}-m^2_{\mathbf{k},\mathbf{k}'}\right)\right] \nonumber \\
&\frac{\mathrm{d}}{\mathrm{d}t}\rho_3(\mathbf{k})\bigg|_\mathrm{e-i}\nonumber \\
&=-\pi n_\mathrm{i}\sum_{\mathbf{k}'}V^2_{\mathbf{k}-\mathbf{k}'}\{\left[\rho_3(\mathbf{k}')-\rho_3(\mathbf{k})\right]\left[\delta(E_{\mathbf{k}}-E_{\mathbf{k}'})+\delta(E_{\mathbf{k}}+E_{\mathbf{k}'})\right. \nonumber \\ &\qquad\qquad\left. +\left(\delta(E_{\mathbf{k}}-E_{\mathbf{k}'})-\delta(E_{\mathbf{k}}+E_{\mathbf{k}'})\right)\left(l^2_{\mathbf{k},\mathbf{k}'}-m^2_{\mathbf{k},\mathbf{k}'}\right)\right] \nonumber \\
&+2\left[\rho_1(\mathbf{k}')-\rho_1(\mathbf{k})\right]\left(\delta(E_{\mathbf{k}}-E_{\mathbf{k}'})-\delta(E_{\mathbf{k}}+E_{\mathbf{k}'})\right)m_{\mathbf{k},\mathbf{k}'}l_{\mathbf{k},\mathbf{k}'}\}\,. 
\end{align}
Here, we introduced
\begin{align}
l_{\mathbf{k},\mathbf{k}'}=u_{\mathbf{k}}u_{\mathbf{k}'}-v_{\mathbf{k}}v_{\mathbf{k}'}\,,\quad m_{\mathbf{k},\mathbf{k}'}=u_{\mathbf{k}}v_{\mathbf{k}'}+v_{\mathbf{k}}u_{\mathbf{k}'}\,, 
\end{align}
with coherence factors
\begin{align}
u_\mathbf{k}=\sqrt{\frac{1}{2}\left(1+\frac{\varepsilon_\mathbf{k}}{E_\mathbf{k}}\right)}\,,\quad v_\mathbf{k}=\frac{\Delta^0_\mathbf{k}}{|\Delta^0_\mathbf{k}|}\sqrt{\frac{1}{2}\left(1-\frac{\varepsilon_\mathbf{k}}{E_\mathbf{k}}\right)}\,, 
\end{align}
and quasiparticle energy
\begin{align}
E_\mathbf{k}=\sqrt{\varepsilon_\mathbf{k}^2+|\Delta^0_\mathbf{k}|^2}\,.
\end{align}
Note that, in our calculations here,  the scattering terms arising from disorder contain the equilibrium coherence factors and quasiparticle energy for simplicity. The latter  are determined by the equilibrium superconducting order parameter $\Delta^0_\mathbf{k}$. This is a good approximation in the weak excitation regime where the non-equilibrium $\Delta_\mathbf{k}(t)$ remains close to its equilibrium value. The delta functions in Eq.~(\ref{eq:scat}) ensure the energy conservation in the electron--impurity scattering processes.

To obtain an expression for the normal state Drude scattering rate, we first study the equations of motion~(\ref{eq:eom_ps}) in the normal state where $T>T_\mathrm{c}$:
\begin{align}
\label{eq:normal_state}
\frac{\mathrm{d}}{\mathrm{d}t}\rho_0(\mathbf{k})&=-e\mathbf{E}\cdot\nabla_\mathbf{k}\rho_3(\mathbf{k})\nonumber \\
&+2\pi n_\mathrm{i}\sum_{\mathbf{k}'}V^2_{\mathbf{k}-\mathbf{k}'}\left[\rho_0(\mathbf{k}')-\rho_0(\mathbf{k})\right]\delta(\varepsilon(\mathbf{k})-\varepsilon(\mathbf{k}'))\,, \nonumber \\
%----------------
\frac{\mathrm{d}}{\mathrm{d}t}\rho_1(\mathbf{k})&=\frac{\mathrm{d}}{\mathrm{d}t}\rho_2(\mathbf{k})=0\,,\nonumber \\
%----------------
\frac{\mathrm{d}}{\mathrm{d}t}\rho_3(\mathbf{k})&=-e\mathbf{E}\cdot\nabla_\mathbf{k}\rho_0(\mathbf{k}) \nonumber \\
&-2\pi n_\mathrm{i}\sum_{\mathbf{k}'}V^2_{\mathbf{k}-\mathbf{k}'}\left[\rho_0(\mathbf{k}')-\rho_0(\mathbf{k})\right]\delta(\varepsilon(\mathbf{k})-\varepsilon(\mathbf{k}'))\,.
\end{align}
The above equations correspond to semiclassical Boltzmann equations~\cite{Kitamura:2015}. As demonstrated in Refs.~\cite{Kitamura:2015,Higgs_dwave_thoery2}, the Drude conductivity, 
\begin{align}
\sigma(\omega)=\frac{n_\mathrm{e} e^2 \tau(k_\mathrm{F})}{m^\star}\frac{1}{1-\mathrm{i}\omega\tau(k_\mathrm{F})}\,,
\end{align}
can be derived from the quantum kinetic equations~(\ref{eq:normal_state}), with electron density $n_\mathrm{e}$, effective mass $m^\star$, Fermi wave vector $k_\mathrm{F}$, and normal state scattering rate
\begin{align}
\frac{1}{\tau(k)}=2\pi n_\mathrm{i}\sum_{\mathbf{k}'}V^2_{\mathbf{k}-\mathbf{k}'}\left(1-\cos\theta_{\mathbf{k},\mathbf{k}'}\right)\delta(\varepsilon(\mathbf{k})-\varepsilon(\mathbf{k}'))\,.
\end{align}
As a result, $\Gamma=\frac{1}{\tau(k_\mathrm{F})}$ corresponds to the  normal-state scattering rate which is experimentally obtained by THz conductivity measurements.

\textbf{Simulation of 2D THz coherent spectra}

For a thin film geometry, the transmitted $E$-field that drives the SC dynamics can be expressed based on Maxwell's equations as follows~\cite{Mootz2020}:
\begin{align}
E(t)=E_0(t)-\frac{\mu_0 c}{2n}J(t)\,,
\label{eq:Etrans}
\end{align}
where $E_0(t)$ is the applied laser electric field, $n$ is the refractive index while
\begin{align}
\label{curr}
J=2\,e\sum_{\mathbf{k}}\nabla_\mathbf{k}\varepsilon(\mathbf{k}) \ \rho_{0}(\mathbf{k})
\end{align}
is the gauge-invariant current. We include the electromagnetic propagation inside the thin SC film in our calculation by self-consistently solving Eq.~(\ref{eq:Etrans}) and the gauge-invariant SC Bloch equations~(\ref{eq:eom_ps}). In order to simulate the measured 2D THz spectra, we adopt the collinear geometry used in the experiment where the SC system is excited by two collinearly propagating THz electric fields. The response of the SC film to these two phase-locked pulses is obtained by calculating the transmitted $E$-field, $E_\mathrm{AB}(t,\tau)$, which depends on both the sampling time $t$ (real time) and the time delay $\tau$ between the two driving electromagnetic fields. The nonlinear differential transmission measured in the experiment is then given by 
\begin{align}
\mathcal{E}_\mathrm{NL}(t,\tau)=E_\mathrm{AB}(t,\tau)-E_\mathrm{A}(t)-E_\mathrm{B}(t,\tau)\,.
\label{eq:Enl}
\end{align}
Here, $E_\mathrm{A}(t)$ ($E_\mathrm{B}(t,\tau)$) is the transmitted electric field induced by pulse A (B).

In this work, we performed simulations that compare  an $s$-wave order parameter, $\Delta_\mathbf{k}=\Delta_0$, with  a $d_{x^2-y^2}$-wave, $\Delta_\mathbf{k}=\Delta_0 (\cos(k_x)-\cos(k_y))/2$, SC order parameter, with $2\Delta_0=1.2$~THz. We considered a tight-binding band dispersion with a square lattice nearest-neighbor structure given by $\varepsilon(\mathbf{k})=-J[\mathrm{cos}(k_x)+\mathrm{cos}(k_y)]+\mu$ with hopping parameter $J=50.0$,~meV and $\mu=0$.  Please note that previous magnetotransport measurement of the superconductivity in nikelate flilms reveals that the superconductivity in nickelates is in the dirty limit~\cite{nickelate_dirtylimit}, so we chose a normal-state scattering rate of $\Gamma=3$~THz to fulfill the dirty-limit superconductivity in nickelates. Note that the simulation results in the main text Fig. 3 do not crucially depend on the strength of $\Gamma$. We self-consistently solved the superconductor Bloch equations~(\ref{eq:eom_ps}), Eq.~(\ref{eq:Etrans}), and the order parameter equation (\ref{gap}) in the time domain for a $400\times 400$ square lattice, using a fourth-order Runge--Kutta method. The SC system is excited with two equal broadband THz electric fields with center frequency $\omega_0=1$~THz comparable to the THz pulses used in the experiment.

\textbf{Data availability.} All relevant data are available on reasonable request from J.W.

\bibliography{scibib}

 \section{Acknowledgements:}
This work was supported by the U.S. Department of Energy, Office of Basic Energy Science, Division of Materials Sciences and Engineering (Ames National Laboratory is operated for the U.S. Department of Energy by Iowa State University under Contract No. DE-AC02-07CH11358). B.C. was supported by the Laboratory Directed Research and Development project, Ames National Laboratory. Work at SIMES (K.L., Y.H.L., B.Y.W., Z.X.S., and H.Y.H.,) was supported by the U.S. Department of Energy, Office of Science, Basic Energy Sciences, Materials Sciences and Engineering Division under Contract No. DE-AC02-76SF00515 and the Gordon and Betty Moore Foundation’s Emergent Phenomena in Quantum Systems Initiative (Grant No. GBMF9072, synthesis equipment).

\section{Author contributions:}
 
B.C., Z.X.S., H.Y.H., and J.W. initiated the project; B.C. and D.C. performed the 2D THz measurements of nickelate films with the help of C.K.H., and L.L.; C.K.H. performed the 2D THz measurements of Nb films with the help of L.L.; M.M., and I.E.P. performed 2D THz calculations; K.L., Y.H.L., B.Y.W., and H.Y.H. developed samples and performed transport characterizations; B.C., J.W., analyzed the spectroscopy data with the help of C.K.H., L.L., and Z.Y.C.; The manuscript was written by B.C., J.W., I.E.P., and M.M. with input from all authors; J.W. supervised the project.

\section{Additional information:} 
 
\textbf{Supplementary Information} accompanies this paper at

\bigskip
 
\textbf{Competing financial interests:} The authors declare no competing financial interests.
 
\bigskip
 
\textbf{Reprints and permission} information is available online at 
\bigskip

\normalsize

\end{document}